\documentclass[twocolumn,showpacs,preprintnumbers,amsmath,amssymb,epsfig,prb]{revtex4}
\usepackage{graphicx}
\begin{document}

\title{Curie temperature and carrier concentration gradients in MBE grown Ga$_{1-x}$Mn$_x$As layers}
\author{A. Koeder\footnote{Electronic-mail: achim.koeder@physik.uni-ulm.de}, S. Frank, W. Schoch, V. Avrutin, W. Limmer, K. Thonke, R. Sauer, and A. Waag}
\affiliation{Abteilung Halbleiterphysik, Universit\"{a}t Ulm, D-89069 Ulm, Germany}
\author{M. Krieger, K. Zuern, and P. Ziemann}
\affiliation{Abteilung Festk\"{o}rperphysik, Universit\"{a}t Ulm, D-89069 Ulm, Germany}
\author{S. Brotzmann and H. Bracht}
\affiliation{Institut f\"{u}r Materialphysik, Universit\"{a}t M\"{u}nster, Wilhelm-Klemm-Strasse 10,
D-48149 M\"{u}nster, Germany}


\begin{abstract}
We report on detailed investigations of the electronic and magnetic properties of
ferromagnetic Ga$_{1-x}$Mn$_x$As layers, which have been fabricated by low-temperature
molecular-beam epitaxy. Superconducting quantum interference device measurements reveal a
decrease of the Curie temperature from the surface to the Ga$_{1-x}$Mn$_x$As/GaAs
interface. While high resolution x-ray diffraction clearly shows a homogeneous Mn
distribution, a pronounced decrease of the carrier concentration from the surface towards
the Ga$_{1-x}$Mn$_x$As/GaAs interface has been found by Raman spectroscopy as well as
electrochemical capacitance-voltage profiling. The gradient in Curie temperature seems to
be a general feature of Ga$_{1-x}$Mn$_x$As layers grown at low-temperature. Possible
explanations are discussed.
\end{abstract}


\maketitle

Over the past years, the field of spin electronics, which aims at the use of the spin of
carriers for electronic devices, is of growing interest. A combination of conventional
semiconductors with magnetism, where both band gap engineering as well as magnetic
engineering could be realized,  would be very desirable. One material under investigation
is the semimagnetic semiconductor Ga$_{1-x}$Mn$_x$As, which can be made ferromagnetic
with Curie temperatures $T_C$ as high as 150K,\cite{Kuannealing} as reported so far. This
ferromagnetism is attributed to be due to the exchange interaction between holes and the
magnetic moment of Manganese , and hence $T_C$ strongly depends on both the Mn content
$x$ and the concentration $p$ of holes:\cite{DietlScience287zenermodell}
\begin{eqnarray}
T_C\propto x \times p^{1/3}
\end{eqnarray}.

The high density of holes in Ga$_{1-x}$Mn$_x$As originates from the incorporation of Mn
itself, which acts as an acceptor on Ga lattice sites and in addition provides the
magnetic moments.

In this work, we have investigated the magnetic and electronic properties of
ferromagnetic Ga$_{1-x}$Mn$_x$As layers by superconducting quantum interference device
(SQUID) measurements, electrochemical capacitance-voltage (ECV) profiling as well as
Raman measurements. A pronounced vertical gradient of the carrier concentration was
found, which consequently leads to a gradient of the ferromagnetic coupling and therefore
to a variation in $T_C$ from the surface towards the Ga$_{1-x}$Mn$_x$As/GaAs interface.
Our results indicate that this obviously is a general aspect of low-temperature
molecular-beam epitaxy (MBE) grown Ga$_{1-x}$Mn$_x$As layers and has to be taken into
account in the analysis of their physical properties.

Ga$_{1-x}$Mn$_x$As thin films were grown by low-temperature MBE at a substrate
temperature of T$_S$ = 250 $^{\circ}$C. For providing As$_4$ a valved As cracker cell was
used in the noncracking mode with a As/Ga beam equivalent pressure ratio of 30, whereas
for Ga and Mn conventional Knudsen cells were used. The growth procedure was as follows:
A 100-nm-thick GaAs buffer layer was grown at high temperature (585 $^{\circ}$C) on a
semi-insulating GaAs(001) substrate, then the sample was cooled down during a growth
break of 45 min to T$_S$ = 250 $^{\circ}$C, and finally the Ga$_{1-x}$Mn$_x$As layer was
grown. Below we will discuss our findings on the basis of the Ga$_{1-x}$Mn$_x$As layer
with $x=5.1\%$, measured by HRXRD, where the effects observed in all samples under
investigation are most pronounced. After the growth the sample was etched by wet chemical
etching to 100 nm and 140 nm etch depth, respectively.The unetched and etched pieces of
the same sample were analyzed by SQUID magnetization measurements, Raman spectroscopy  as
well as ECV profiling.

\begin{figure}
\includegraphics[width=0.5\textwidth]{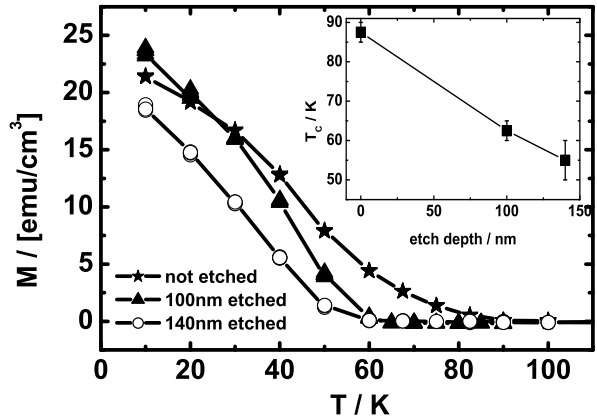}
\caption{\label{MvsT} Magnetization curves taken from the as-grown and from the etched
samples. The inset shows the decrease of $T_C$ versus etch depth.}
\end{figure}

SQUID magnetization measurements show a decrease in $T_C$ from 90 K in the as-grown
sample to 55 K in the sample with 140 nm etch depth (Fig.~\ref{MvsT}).This corroborates
results obtained from ferromagnetic resonance measurements,\cite{Goennenwein} which have
already pointed to the fact that there is a gradient in the magnetic properties. The
inset of Fig.~\ref{MvsT} shows the decrease in measured T$_C$ versus etch depth.

According to Eq.~(1), a change in Curie temperature can be caused either by a gradient in
Mn content $x$ or by a gradient in the carrier concentration $p$.

\begin{figure}
\includegraphics[width=0.5\textwidth]{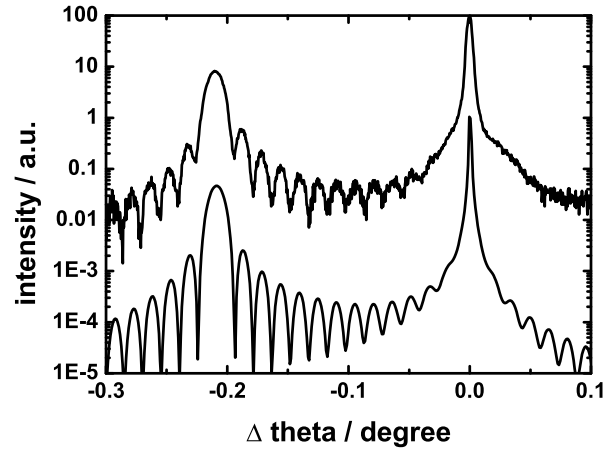}
\caption{\label{HRXRD} HRXRD measurement (upper curve) and simulation (lower curve) of
the as-grown Ga$_{0.949}$Mn$_{0.051}$As layer.}
\end{figure}

To check for a gradient in Mn distribution, we performed HRXRD measurements. A Mn
gradient should affect the shape of the HRXRD curve. In Fig.~\ref{HRXRD}, the measured
HRXRD spectrum of the as-grown sample is shown as well as a simulation, assuming a
homogeneous Mn content $x=5.1\%$. The position of the Ga$_{1-x}$Mn$_x$As peak as well as
the full width half maximum (FWHM) of the measurement agrees nearly perfectly with the
simulation. Therefore, a gradient in Mn content as the origin of the observed decrease in
$T_C$ can be ruled out. The Ga$_{1-x}$Mn$_x$As layers are homogeneous in terms of lattice
constant, and are of high quality, as can also be seen by the pronounced thickness
fringes, indicating a sharp Ga$_{1-x}$Mn$_x$As/GaAs interface.

\begin{figure}
\includegraphics[width=0.5\textwidth]{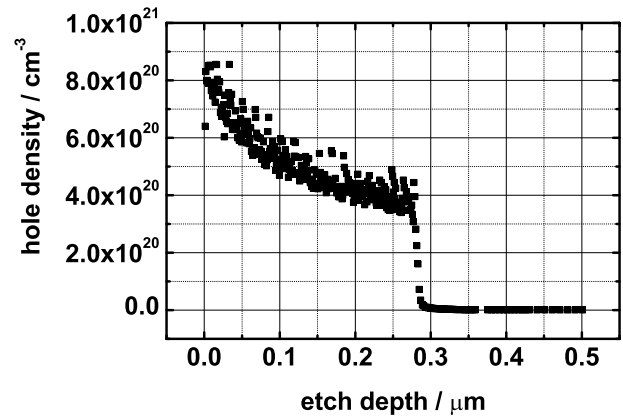}
\caption{\label{ecv} Electrochemical capacitance-voltage profiling of the as grown
Ga$_{0.949}$Mn$_{0.051}$As layer.}
\end{figure}

To check for a vertical gradient in hole concentration $p$ across the Ga$_{1-x}$Mn$_x$As
layer, ECV profiling is a suitable method\cite{YuECV} overcoming the complications that
arise from the anomalous Hall effect which affects standard transport studies of carrier
concentration in conducting ferromagnetic materials.\cite{Matsukuratransport} ECV
profiling measurements using a BioRad PN4400 with a 0.2M NaOH:EDTA solution as the
electrolyte were performed on an as-grown piece of the sample, which was used for Raman
and SQUID measurements. Indeed, the ECV profile also shows a decrease in hole
concentration by a factor of 2 from $8 \times 10^{20}$cm$^{-3}$ to $4 \times
10^{20}$cm$^{-3}$ from the surface to the Ga$_{1-x}$Mn$_x$As/GaAs interface
(Fig.~\ref{ecv}).

\begin{figure}
\includegraphics[width=0.4\textwidth]{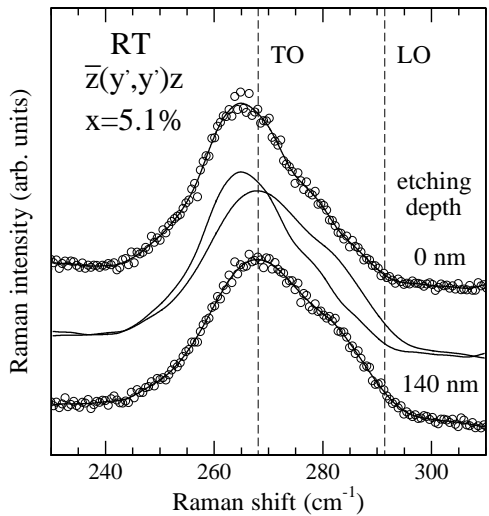}
\caption{\label{raman} Raman spectra (open circles) recorded from the as-grown sample and
from the 140-nm-etched sample. For a better comparison of the measured spectra the
corresponding spline curves (solid lines) are plotted separately. The dashed lines mark
the TO- and LO-phonon positions in the GaAs substrate.}
\end{figure}

To confirm the result of ECV profiling we have performed Raman spectroscopy measurements.
The Raman signals of the unetched and the 140-nm-etched sample are compared in
Fig.~\ref{raman}. As discussed in detail in Ref. 4 the high hole concentration in
Ga$_{1-x}$Mn$_x$As leads to the formation of a coupled mode of the longitudinal optical
(LO) phonon and the hole plasmon. With increasing hole concentration the frequency of
this coupled mode shifts from the frequency of the LO phonon to that of the transverse
optical (TO) phonon. Figure~\ref{raman} clearly reveals that the coupled mode in the
etched sample is shifted towards the LO-phonon frequency if compared to the unetched
sample, indicating a reduced hole concentration.

As a result, we have detected a vertical gradient in the hole concentration, leading to a
vertical gradient in the Curie temperature. We suggest that this gradient is generally
occurring in Ga$_{1-x}$Mn$_x$As layers grown by LT-MBE, as will become clear from the
discussion below, and has to be taken into account when measuring the properties of such
samples.

Meanwhile it is well known that post-growth annealing of Ga$_{1-x}$Mn$_x$As at
temperatures near the growth temperature can change the magnetic behaviour of these
layers .\cite{Kuannealing}$^,$\cite{Yuinterstitials}$^-$\cite{Foxon} The origin of this
annealing effect is obviously due to changes in the point defect structure of the low
temperature grown Ga$_{1-x}$Mn$_x$As. E.g., diffusion of As antisites\cite{Potashnik} and
the rearrangement of Mn at interstitial sites\cite{Yuinterstitials} or a combination of
both are under discussion. The main point is that these effects are observed during
annealing at temperatures near or even below the Ga$_{1-x}$Mn$_x$As growth temperature.

If annealing near or even at the growth temperature changes the defect structure, then
this should also happen during growth. Whether this annealing leads to a higher or lower
Curie temperature depends, however, on the particular nature of point defects in the
structure. The exact microscopic processes are very complex and still unclear.

In addition to this annealing effect, a change of surface temperature during growth can
be also a source of vertical gradients. When Mn is incorporated into GaAs, the effective
bandgap E$_g$ shrinks \cite{VanEsch}, and the free carrier concentration goes up
($10^{20}$cm$^{-3}$ or higher). As a consequence, the absorption of infrared light coming
from the hot effusion cells in the MBE system increases and thus the surface temperature
gradually rises. An increased growth temperature again can change the point defect
structure of the Ga$_{1-x}$Mn$_x$As substantially, which in turn affects the magnetic and
electrical properties.

The observed vertical gradient in T$_C$ can also explain the unusual magnetization curves
M(T) of as-grown samples measured by our group (Fig.~\ref{MvsT}), but also by
others.\cite{Kuannealing}$^,$\cite{Potashnik}$^,$\cite{VanEsch}$^,$\cite{Ohno2001SSC} The
M(T) curves of Ga$_{1-x}$Mn$_x$As layers usually do not show a simple Curie-Weiss
behaviour. This can be explained by the superposition of various M(T) curves from parts
of the layer with different Curie temperature. A detailed investigation of magnetization
curves will be published elsewhere.

In conclusion we have shown that as-grown Ga$_{1-x}$Mn$_x$As layers potentially show a
vertical gradient in carrier concentration and therefore in their magnetic properties. We
suggest that this is due to the intrinsic high point defect density, which leads to high
diffusion constants of point defects already at growth temperatures. Even though the
vertical gradient in Curie temperature depends on the particular growth process used, it
is suggested to be a general and intrinsic property of low-temperature Ga$_{1-x}$Mn$_x$As
layers, which can lead to a severe misinterpretation of the data if not taken into
account properly.

The authors acknowledge financial support by the Deutsche Forschungsgemeinschaft, DFG Wa
860/4-1.

\end{document}